\documentclass[12pt,preprint]{aastex}
\usepackage[]{natbib}
\usepackage{graphics}
\usepackage{amssymb}
\usepackage{amsmath}
\def\sun{$_{\mathrm \odot}$}
\def\deg{$^{\mathrm o}$~}
\def\ergs{erg~s$^{-1}$~}
\def\kms{km~s$^{-1}$~}
\slugcomment{} \shorttitle{two extreme double-peaked line
emitters} \shortauthors{Wang et al.}
\slugcomment{ApJL accepted}

\begin{document}

\title{Two extreme double-peaked line emitters in the Sloan Digital Sky Survey}
\author{T.G. Wang, X. B. Dong, X.G. Zhang, H.Y. Zhou, J. X. Wang and Y.J. Lu}
\altaffilmark{Center for Astrophysics, University of Science and Technology of China, Hefei, Anhui, 230026, P.R.China (Email:twang@ustc.edu.cn)}

\begin{abstract}
Double-peaked Balmer lines have been observed in about 150 AGNs
and were interpreted preferably as emission from relativistic
accretion disks. In this paper, we report the discovery of extreme
double-peaked lines in SDSS J0942+0900 and SDSS J1417+6141. The FWHM of
the H$\alpha$ line $\simeq$ 40,600 km~s$^{-1}$ in the first object is
almost twice as large as the broadest one previously known. By
comparing the line profile with accretion disk models, we find
that most of the line flux is emitted from a narrow annulus around  
100$r_g$ in SDSS J0942+0900, and from a disk of radii between 100
and 400$r_g$ in SDSS J1417+6141. This is the first time that an
accretion disk at radii below 100$r_g$ can be directly probed
through optical emission lines. A certain asymmetry in the disk is
required for both objects. Another much weaker broad H$\alpha$
component (W$_\lambda\sim$20\AA, and FWHM $\sim$4000~km~s$^{-1}$)
is also detected in both objects. Both objects show weak radio
emission and strong high-ionization narrow lines.

\end{abstract}


\section{Introduction}

Double-peaked low-ionization broad lines are detected in over 150
AGNs now. Most of these objects were discovered either during the
spectroscopic survey of radio loud AGNs (Eracleous \& Halpern 2003, 
hereafter EH03) or during the Sloan Digital Sky Survey (SDSS, Strateva 
et al.  2003, hereafter S03). The incidence of such lines is 
particularly high in radio loud AGNs ($\sim$20\%; EH03), and in 
Low Ionization Nuclear Emission Line Regions (LINERs) with broad 
emission lines (Shields et al. 2000 and references therein). 

The double-peaked line profile was proposed to be the
characteristic feature of lines emitted from an accretion disk
(Chen \& Halpern  1989, hereafter CH89), from a binary accreting
black hole system (Gaskell 1988), or biconical outflows (Zheng et
al. 1991). EH03 and S03 argued that the observed properties of
double-peaked line emitters favor the disk line model (see also
Halpern \& Filippenko 1988). By applying relativistic disk models
to the double-peaked line profiles, they suggested that inner disk
radii are around 200$-$800$r_g$ ($r_g\equiv GM/c^2$) and outer
radii larger than 2,000$r_g$, and that significant deviation from
circular, relativistic Keperian disk is required in about
40$-$60\% of the objects (S03; EH03).

Relativistic disk lines have also been detected in X-ray band
(e.g., Fe K$\alpha$ line; Fabian et al. 2000). These lines are
produced at much smaller radii($\le 10r_g$). 
X-ray emission lines from the disk at larger radii might also have
been detected, but cannot be confirmed up to now due to the
limitation of the spectral resolution and sensitivity of current
X-ray missions (e.g., Lu \& Wang 2000). 
In this paper, we report
the discovery of two extreme double-peaked line emitters SDSS
J094215.1+090015 (SDSS J0942+0900; z=0.21262) and SDSS
J141742.95+614152 (SDSS J1417+6141; z=0.119) in the Sloan Digital
Sky Survey (SDSS: York et al. 2000) data release 3 (DR3: Abazajian
et al. 2005). 

\section{Observation and data reduction}

\noindent {\bf SDSS J0942+0900:} It was noticed for its anomalous
emission lines during our systematic modelling of the continuum
and emission lines of SDSS spectra classified as QSO or galaxies
by the SDSS pipeline. The PSF magnitudes (AB) are 17.921, 17.387,
16.946, 16.688, 16.552, in $u$, $g$, $r$, $i$ and $z$,
respectively.  It was detected in the 2 Micron All Sky Survey
(2MASS) with total magnitudes of 15.48$\pm$0.06, 14.98$\pm$0.09
and 14.30$\pm$0.06 in J, H and K$_s$ bands, respectively, and in
the FIRST survey (White et al. 1997) with a radio flux of
1.93$\pm$0.15 mJy at 20 cm and a compact radio morphology.
Following Ivezi\'c et al (2002), we find the ratio of
radio-to-optical flux density $R_i \equiv \log
(f_{20cm}/f_i)=0.58$, where $f_i$ and $f_{20cm}$ are the flux
densities at \textit{i}-band and 20~cm, respectively. Thus this
quasar is radio quiet. The spectral energy distribution (SED; $\nu
f_\nu$) peaks at the near-infrared, which is also typical for
double-peaked line objects (EH03).

The optical spectrum of SDSS J0942+0900 in the source rest frame,
after correction for Galactic reddening of
E(B$-$V)=0.031 mag (Schlegel et al. 1998), is presented in
Figure~1. The most surprising characteristic is the presence of
very broad, double-peaked Balmer lines. The line profile of broad
H$\alpha$ extends over a wavelength range of about 900\AA~ in the
source rest frame. The H$\beta$ line displays a similar profile, but
its blue peak is blended with the expected red peak of the H$\gamma$
line. Similar profiles may also be present in higher order
Balmer lines, but it is difficult to identify them due to their
weakness and the effect of line-blending.

To extract the H$\alpha$ line profile, we fit the continuum in the
wavelength range of 5400$-$7500\AA~ with a power-law. Apparent
emission lines are masked during the fit. This yields a spectral
index of $\alpha=1.03$ ($f_\nu\propto \nu^{-\alpha}$).
Extrapolating the power-law to short wavelengths predicts a
continuum higher than the observed one, suggesting the presence of
reddening or intrinsic steepening of spectrum in the UV band.
Figure 2 shows the line profile of H$\alpha$ (left panel). The profile 
is apparently skewed to the red, and the blue peak is higher than the 
red one. These features are signatures of lines originating from 
a relativistic disk. 

A circular disk model is considered first. Before making a
detailed fit, we can estimate approximately the inner, outer radii
and the inclination by using the two peak positions, the maximum
extension of the red and blue wings, following the expression of
Doppler factor for circular disk model given by CH89. The two
peaks at 6265 and 7053\AA~in the source rest frame suggest an
outer radius of around 110$r_g$, and an inclination angle
$i\sim38$\deg ( $i=0$\deg for a face-on disk). The maximum
extensions of blue and red wings at 6191\AA~and 7119\AA~ suggest
an inner radius of around 90$r_g$. Thus most of the line flux is
produced in a narrow annulus of radii around 100 $r_g$.

A detailed fit to the H$\alpha$ profile with a circular,
relativistic Keplerian disk model (CH89) with a power-law
distribution of emissivity ($F(r)\propto r^{-\beta}$) yields
parameters: $r_{in}=98\;r_g$, $r_{out}=227\;r_g$, $i=41$\deg,
$\sigma=1287$\kms, $q=4.6$. Narrow lines ([OI], [NII], [SII] and
narrow H$\alpha$) were masked during the fit.  However, this fit 
does not reproduce either the right position of the red peak or 
the relative height of the two peaks as displayed in Fig 2, 
which is a strong
indication for the deviation from axi-symmetry of the disk, in
emissivity or/and in kinematics. Several types of asymmetry, such
as an elliptical disk, presence of hot spots or a spiral arm may
reproduce the line profile. In order to compare parameters for
this object with those presented in EH03 and S03, elliptical disk
models (Eracleous et al. 1995) are then tried. The profile can be
reasonably well fitted using this model. The derived disk
parameters are: $r_{in}=62$$r_g$, $r_{out}=87$$r_g$, $i=35$\deg,
$\sigma=1335$ \kms, $q=2.7$, $e=0.25$ and $\phi_0=18$\deg. An
additional symmetric broad component of FWHM $\sim$ 3300
km~s$^{-1}$ is required to reproduce the whole H$\alpha$ line. The
equivalent width (EW) of this component is 21\AA.

Because the profile of broad H$\beta$ is similar to that of
H$\alpha$ and blends with H$\gamma$ line, we simply rescale the
best fitted H$\alpha$ model by a factor of 0.45 in the flux
density to match the H$\beta$ line profile. This yields
H$\alpha$/H$\beta\simeq2.9$ for the double-peaked component, which
is typical of Seyfert galaxies and quasars as well as
double-peaked line emitters.

The fluxes of the narrow lines in the wavelength ranges covering the
double-peaked H$\alpha$ and H$\beta$ lines are measured after the
best fitting model is subtracted.  At other wavelengths, we use the
local `continuum' that includes weak higher order double-peaked
Balmer lines. Narrow lines are modelled using Gaussian functions.
Two components are required for [OIII] $\lambda$4959\AA~ and
5007\AA~ separately. The narrow line spectrum is typical for
Seyfert galaxies with [OIII]$\lambda 5007$/H$\beta^{n}\simeq$ 11,
and [NII]$\lambda$6563/H$\alpha^n=$0.25.

\noindent {\bf SDSS J1417+6141:} It is also discovered
serendipitously. The Petrosian magnitudes (AB) are 19.153, 17.882,
16.997, 16.374, 16.23 mag in $u$, $g$, $r$, $i$, $z$,
respectively. It was detected in the 2MASS survey with total
magnitudes 14.78$\pm$0.10, 13.87$\pm$0.10, and 13.39$\pm$0.12 in
J, H, and $K_s$ bands, and in the NVSS survey with a radio flux at
21cm of 7.4$\pm$0.5 mJy. Combining of the flux at 21cm with the
non-simultaneous radio flux at 92 cm indicates a steep radio
spectrum with $\alpha=0.5-0.6$ (Rengelink et al. 1997). It is
also radio quiet. 

The SDSS spectrum is dominated by starlight (Fig 1), which is
subtracted following Dong et al. (2005). The H$\alpha$ line
profile is shown in Fig 2. The FWHM of the broad H$\alpha$ line is
around 26,000 km~s$^{-1}$. The line profile is very different from
that of SDSS J0942+0900. Despite of their large separation, the
two peaks are approximately at symmetric positions. Furthermore,
the red peak is (slightly) higher than the blue one, which means
that the line cannot be produced in a homogenous circular disk.
The fit using a circular disk model is poor shown in
Figure 2 (right panel), with the following parameters:
$r_{in}=250~r_g$, $r_{out}=260r_g$, $i=36$\deg, $\sigma=4200$ \kms,
and $q=0.6$. An elliptical disk model can fit the observed line
profile satisfactorily, with parameters being $r_{in}=93~r_g$,
$r_{out}=396~r_g$, $i=40$\deg, $\sigma=1681$ \kms, $q=2.0$,
$e=0.11$ and $\phi_0=57$\deg. An additional broad component of
FWHM $\sim$ 5900 km~s$^{-1}$ is also required in the fit. The EW
of this component is $\sim$20\AA~ in the source rest frame. The
H$\beta$ flux is estimated using the model of H$\alpha$ line in
the same way as for SDSS J0942+0900. The H$\alpha$/H$\beta$ ratio
is around 5.

\section{Discussion}

\subsection{Comparison with other double-peaked line emitters}

S03 collected a large sample of 116 double-peaked line emitters
from the SDSS quasar sample. They found that the H$\alpha$ lines
in these double-peaked line emitters, with a median FWHM around
8,000 km~s$^{-1}$, are much broader than that of quasars in their
parent sample (see also EH03). The line width of SDSS J0942+0900
is almost twice the maximum in their sample (SDSS J1014+0006:
FWHM=20,700 km~s$^{-1}$).  The H$\alpha$ line in SDSS J1417+6141 is 
the second broadest. Comparing with relativistic disk models, the 
bulk of H$\alpha$ line is produced in a narrow
annulus of radii less than 100$r_g$ in SDSS J0942+0900 and in a
disk of inner radius around 90$r_g$ and outer radius 400$r_g$ in
SDSS J1417+6141. Both inner and outer radii are extremely small
compared with those of objects in S03. A certain asymmetry is
required for both objects as for about half (60\% in S03 and 40\%
in EH03) double-peaked line emitters in S03 and EH03. Under the
prescription of elliptical disk models, the ellipticities of both
objects are typical for double-peaked line emitters. The optical
luminosities [$\nu L_\nu(5100{\mathrm \AA})=1.9\times
10^{44}$\ergs for SDSS J0942+0900, and 1.5$\times 10^{43}$\ergs
for SDSS J1417+6141], and the Balmer decrements are well within the
range for double-peaked line emitters in the SDSS sample.

By comparing the UV line profiles of Arp 102B with those of the 
Balmer lines, Halpern
et al. (1996) found that the contribution of a disk component to
CIII], CIV, and Ly$\alpha$ emission is negligible, suggesting that
high-ionization lines are produced in a different region
; and that an additional broad, non-disk
component is required to fit the H$\alpha$ and H$\beta$ line
profile in Arp 102B. But it is hard to distinguish an additional
broad component from the asymmetrical disk model because of the
closeness of the two peaks and their blending with narrow lines.
In SDSS J0942+0900 and SDSS J1417+6141, the separation between the
two peaks of the H$\alpha$ line is much larger, which makes it
easier to identify any additional component. In both objects, the
normal broad component exists apparently albeit weak. With FWHM
being around 3300$-$6000 km~s$^{-1}$, this component cannot be the
broad wing of the narrow lines, and should come from a different
line-emitting region (e.g., the ionized wind). The EW of this
component is only $\simeq$ 20\AA~ (rest frame). 


Since only two are known, extreme double-peaked line emitters are
certainly very rare. However, we note that such objects might
be not as rare as they appear since the detection of such objects
is technically difficult. First, such a line can only be detected
in spectra with relative high S/N ratio because of its low height.
Since the H$\beta$ line is a factor of three weaker, it is even more
difficult to find them if H$\alpha$ is shifted out of the
spectral coverage. Second, the SDSS pipeline is not designed to
detect a line as broad as this. In fact, the very broad lines in
both SDSS J0942+0900 and SDSS J1417+6142 were taken as part of the
continuum by the SDSS pipeline, and only the narrow cores of the lines
were measured. Selections based on the line parameters given by
the SDSS pipeline will miss such objects. A further problem is
that when the contribution of a stellar continuum is significant,
any weak broad peaks may be mixed with stellar feature. Proper
subtraction of stellar light with the presence of such broad lines
is challenging. Combining all these factors together, there may be
more such objects still to be sought out in the SDSS spectroscopic
sample.

\subsection{The origin of very wide double-peaked lines}

Historically, double-peaked lines were also interpreted as
emission from a binary black hole system, in which each hole
possesses its own BLR. For the wide separation of the peaks in
SDSS J0942+0900, there are additional problems for the model
besides those listed by EH03. First, the profile is not consistent
with the combination of two broad emission lines, which should
show two well defined peaks. Second, the very wide separation of
the two peaks requires the separation of the binary to be order of
100$r_g$, and the size of each BLR should be even smaller. Such
a small BLR would be unprecedented. Furthermore, equal mass binary
of black holes at this separation will merge on a time scale of
$\frac{5}{256}\frac{a_0^4} {\mu M^2}$=10$^4$ ($\frac{M}
{10^8M_\odot})^2$ yrs due to gravitational radiation (Misner,
Thorne \& Wheeler 1973), which makes the probability to see such
objects extremely low.

If the line is emitted from a disk, what makes the line-emitting region so
small? From the consideration of energy balance, CH89 argued
that double-peaked lines originate from a truncated disk, with an
outer thin disk and an inner thick, Low Radiative Efficiency
Accretion Flow (LRAF), presumably with a very low mass accretion
rate (Quataert et al. 1999) and illuminating the outer thin disk.
Within this scenario, we might expect that the truncation radius
depends on the mass accretion rate: at very low accretion rate,
the truncation radius is large and the disk is neutral; as the rate
increases, the truncation radius decreases but the thin disk is
still not ionized; but if the accretion rate increases further, the
inner part of thin disk becomes ionized, which sets the smallest possible
disk radius that can produce significant Balmer line emission
even if the truncation radius shifts inwards. Hence there may
exist a sequence of objects from very low-luminosity AGNs 
with a narrow line profile,
to normal double-peaked line emitters, and
to extremely broad double-peaked line emitters, depending on one 
intrinsic parameter, the accretion rate. SDSS J0942+0900 may
be a fortuitous object that at the right accretion rate,
possessing a disk with a small truncation radius yet partially
ionized.

With a truncation radius of order 50$-$100$r_g$, the thermal emission 
from the disk turns down in the optical or near UV. The spectrum 
steepening in the near UV in SDSS J0942+0900 may be an indication of 
this. 
A further prediction of the
truncated disk model is an iron K line component with a profile
similar to that of the optical H$\alpha$ line and with an EW of a few
tens of eV for the fraction of X-rays intercepted by the disk as
estimated in CH89. The radiative efficiency of such a disk in the
optical/UV is on the order of 0.01, and the presence of an LRAF in these
objects requires the accretion rate considerably lower than the
Edington accretion rate. Thus, to explain the optical luminosity
of 1.9$\times 10^{44}$ ergs~s$^{-1}$ in SDSS J0942+0900, as well
as many other double-peaked line emitters in the SDSS, a black
hole mass of at least 10$^8$ M\sun~ is required.

However, local viscous heating seems sufficient to power the
H$\alpha$ emission in SDSS J0942+0900. We carry out an analysis similar
to that presented in CH89 as follows. The soft X-ray (0.2$-$2 keV)
flux of this object is
1.94$\times$10$^{-13}$~erg~cm$^{-2}$~s$^{-1}$ detected by ROSAT
(WGACAT). Thus its H$\alpha$ to soft X-ray flux ratio is similar
to Arp 102B (CH89), while its inner disk radius is a factor of 5
smaller. For Arp 102B , CH89 found that the H$\alpha$ luminosity
is similar to the local heating rate.  Since the available local heating
rate is proportional to $(r/r_g)^{-1}$,  it
exceeds the H$\alpha$ line luminosity by a factor of $5$ in SDSS
J0942+0900. Thus an inner LRAF as an external illuminating source
is not required.

In Seyfert galaxies and quasars that were targeted for reverberation 
mapping, an
empirical relation between the BLR size and the optical luminosity
was identified (Kaspi et al. 2000). Following their relation, a
size of BLR of 1.3$\times10^{17}$~cm is predicted for SDSS
J0942+0900. Combining this with the mean disk radius, an
extraordinarily large mass of central black hole, $\sim 10^{10}$
M\sun~, is inferred. According to the Magorrian relation
(Magorrian et al. 1998), however, if the black hole is so massive
, the luminosity of the spheroid of its host galaxy would
be impossibly higher than $10^{11}$L\sun, which is $\sim$5 times
larger than the total optical luminosity of SDSS J0942+0900.
Actually, the stellar contribution is likely at the 10\%
level, in the light of the fact that the stellar absorption lines
are weak in this object. A lower black hole mass of order
$10^8$~M$_\odot$ can be estimated from the [OIII] width (440
km~s$^{-1}$ in FWHM) following Nelson (2000), albeit with a large
uncertainty. Therefore, we suggest that the empirical relation
between the BLR size and the optical luminosity does not hold for
double-peaked line emitters.

A lot of work still needs to be done to reveal the nature of the
mysterious extremely broad lines. Constructing an efficient
algorithm to find the number of such objects in the SDSS would
help determine the frequency of such widely separated double-peaked line
emitters. Since the emission line region is much closer to the
black hole, the line profiles should be variable on shorter time
scale than other massive double-peaked line emitters (e.g.,
3C390.3) unless the black hole is extraordinary.
Monitoring the variations of the line intensity and profile would
allow us to determine the origin of the asymmetry, and may give
some constraints on our current understanding of the accretion
disk structure and its evolution. The stellar velocity dispersion
of the host galaxy can be obtained with high S/N ratio and high
spectral resolution observations, yielding an estimate of the
black hole mass, which may help us to constrain the accretion
rate. UV and X-ray observations should allow a further test for the
truncated disk models.

\acknowledgements We thank the referee Dr. Eracleous for many
useful suggestions and help in English. This work is supported by 
Chinese NSF grant NSF-10233030, the Bairen Project of CAS.
Funding for the creation and the distribution of the SDSS Archive 
has been provided by the Alfred P. Sloan Foundation, the Participating
Institutions, the National Aeronautics and Space Administration,
the National Science Foundation, the U.S. Department of Energy,
the Japanese Monbukagakusho, and the Max Planck Society. The SDSS
is managed by the Astrophysical Research Consortium (ARC) for the
Participating Institutions. The Participating Institutions are The
University of Chicago, Fermilab, the Institute for Advanced Study,
the Japan Participation Group, The Johns Hopkins University, Los
Alamos National Laboratory, the Max-Planck-Institute for Astronomy
(MPIA), the Max-Planck-Institute for Astrophysics (MPA), New
Mexico State University, Princeton University, the United States
Naval Observatory, and the University of Washington.

\clearpage

\begin{figure}
\epsscale{1.1}
\plottwo{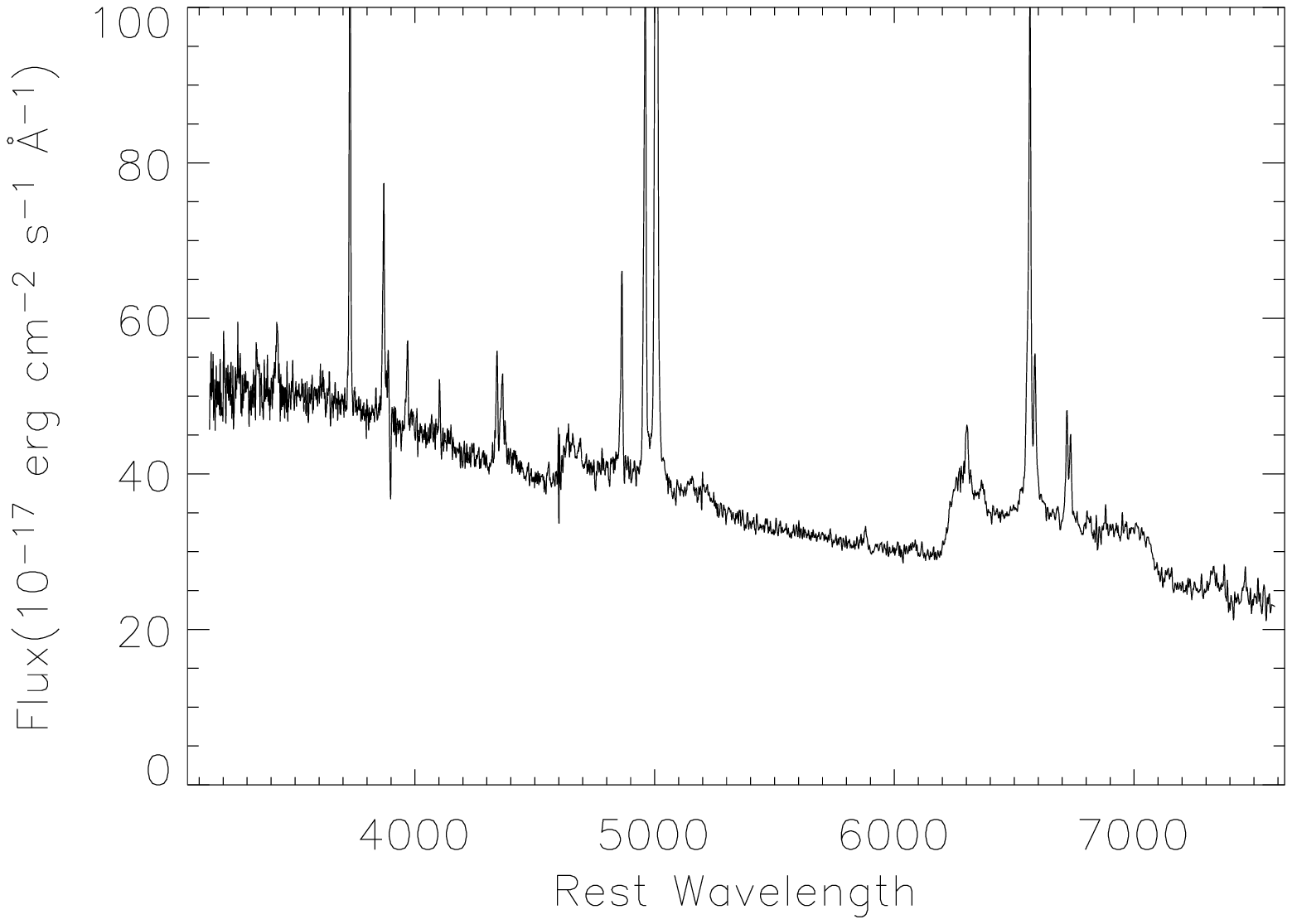}{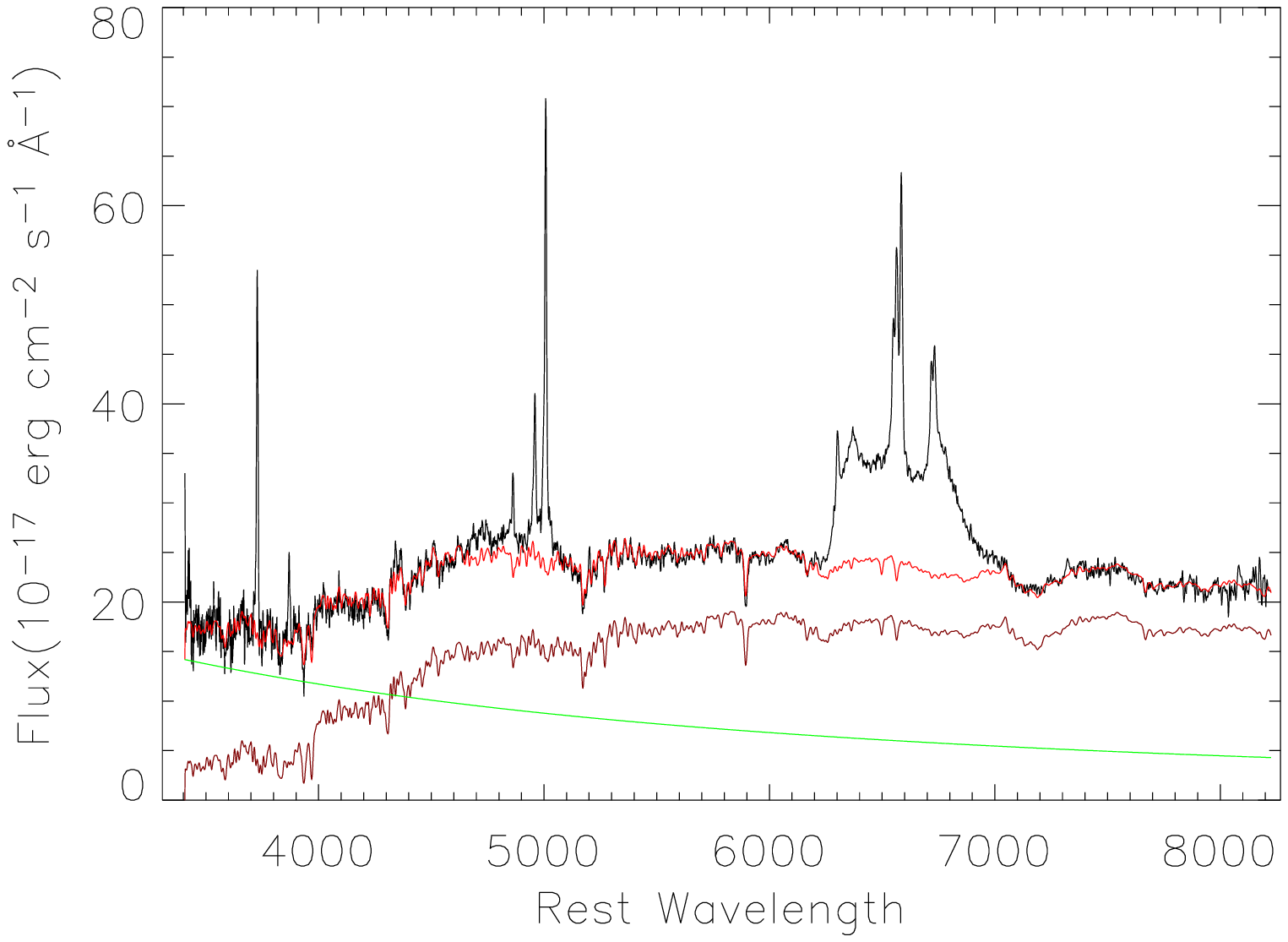}
\caption{The SDSS spectra of SDSS J0942+0900 (left panel) and SDSS
J1417+6141 (right panel), corrected for the Galactic reddening.
The Balmer lines are very broad and double-peaked. The starlight and
nuclear continuum are also shown for SDSS J1417+6141.
\label{fig1}}
\end{figure}

\clearpage

\begin{figure}
\epsscale{1.0}
\plottwo{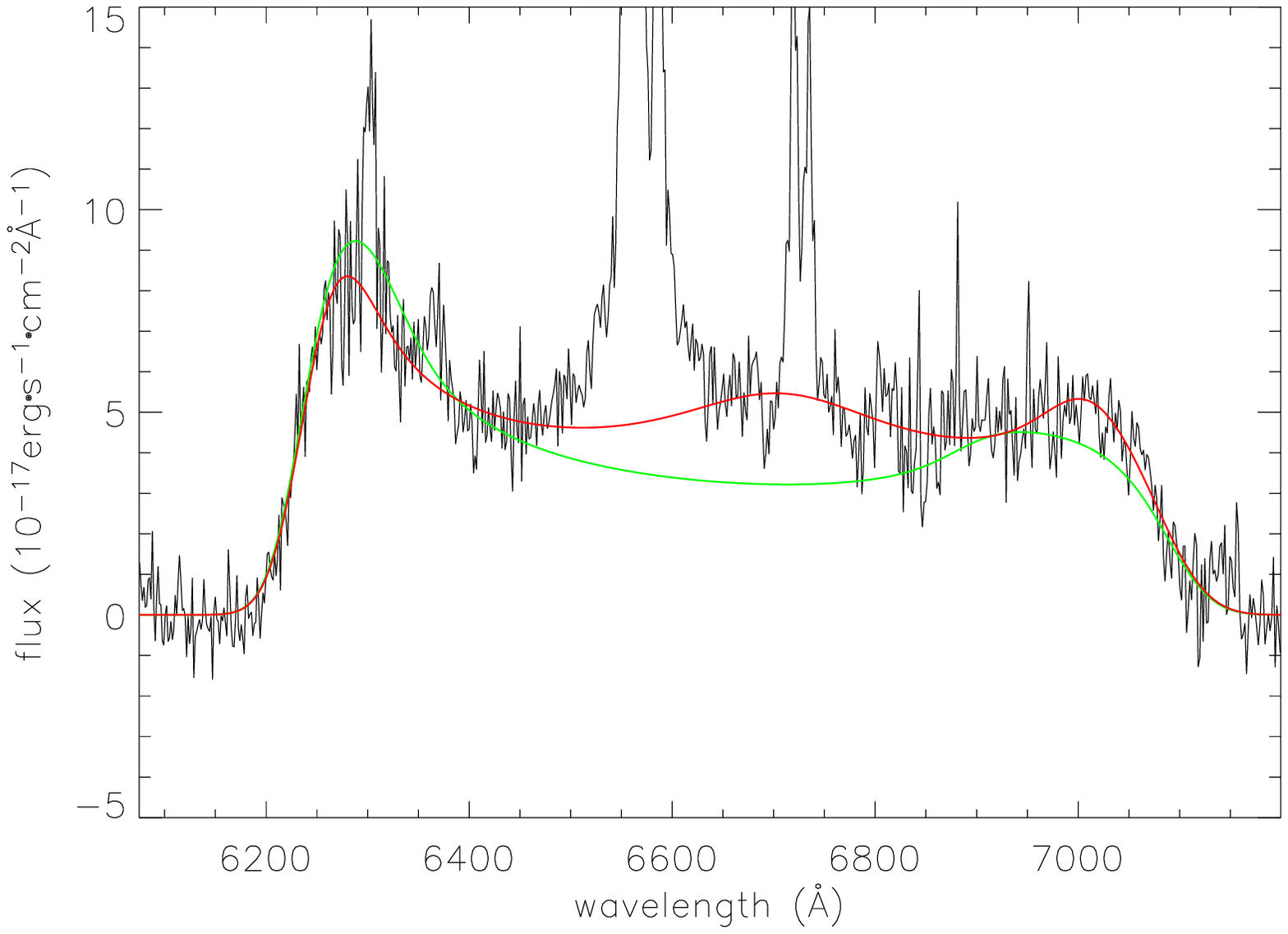}{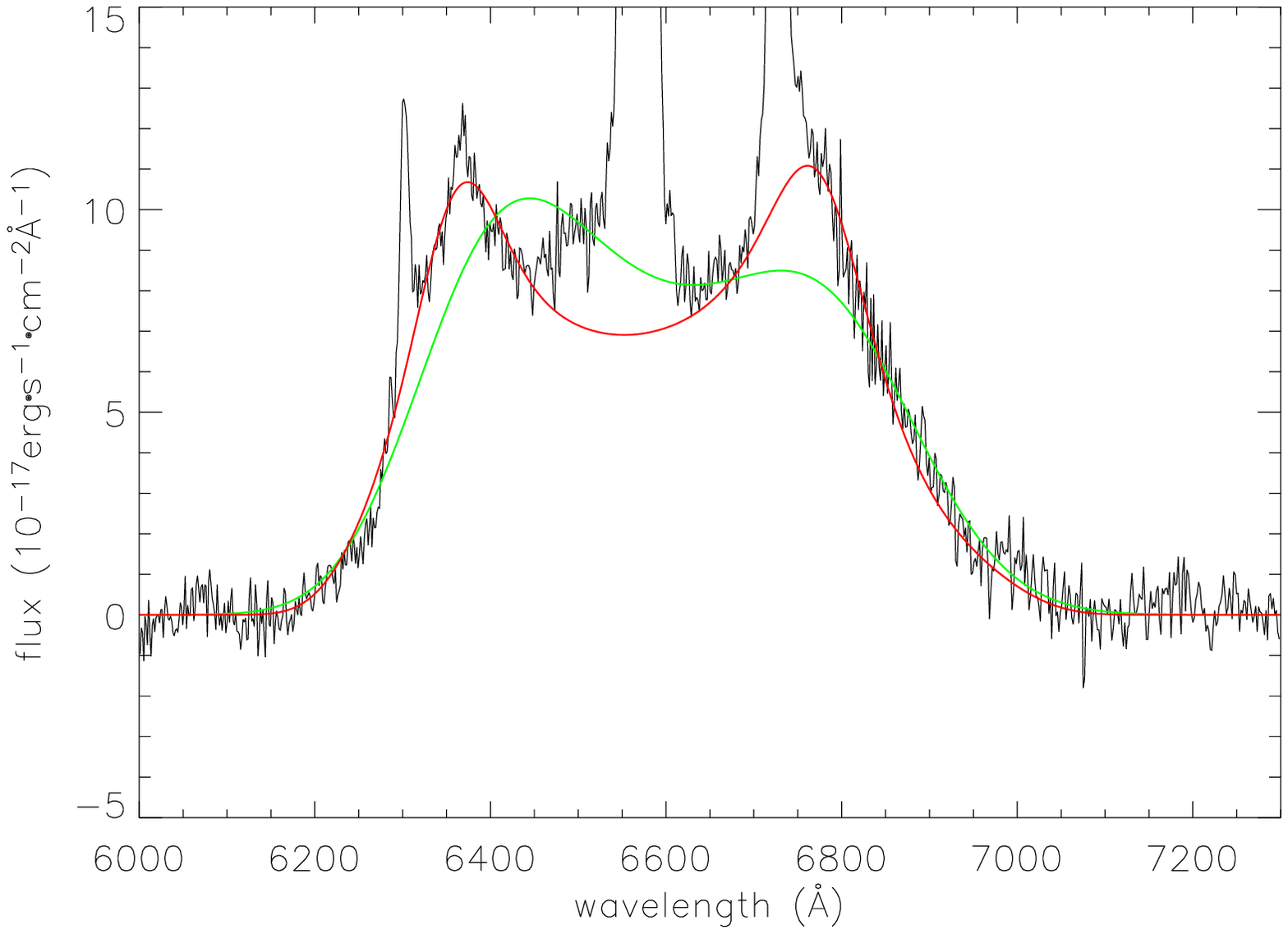}
\caption{The observed H$\alpha$ line profiles and the best fitting
relativistic disk models (red: elliptical disk; green: circular
disk) to the H$\alpha$ lines for SDSS J0942+0900 (left panel) and SDSS
J1417+6141 (right panel). Note that the additional broad H$\alpha$
component in each fit is not displayed.}
\end{figure}

\end{document}